\documentclass[12pt,reqno]{article}
\usepackage{amsmath, amsthm, graphicx,amsfonts, amssymb,color}
\usepackage{bm}
\usepackage{hyperref}
\RequirePackage{amsthm,amsmath,amsfonts,amssymb,mathtools}
\RequirePackage[numbers]{natbib}

\usepackage{bm}
\usepackage{hyperref}
\usepackage{mathrsfs}

\newtheorem{thm}{Theorem}[section]
\newtheorem{lemma}[thm]{Lemma}

\newtheorem{corol}[thm]{Corollary}
\newtheorem{propos}[thm]{Proposition}
\newtheorem{rema}{Remark}[section]
\newtheorem{defini}[thm]{Definition}

\def\bd{\begin{defini}}
	\def\ed{\end{defini}}
\def\bp{\begin{propos}}
	\def\ep{\end{propos}}
\def\bt{\begin{thm}}
	\def\et{\end{thm}}
\def\bco{\begin{corol}}
	\def\eco{\end{corol}}
\def\bl{\begin{lemma}}
	\def\el{\end{lemma}}
\def\br{\begin{rema}}
	\def\er{\end{rema}}
\def\be{\begin{equation}}
	\def\ee{\end{equation}}
\def\ba{\begin{array}}
	\def\ea{\end{array}}
\def\bena{\begin{eqnarray}}
	\def\eena{\end{eqnarray}}

\def\P{{\mathbb P}}

\def\R{{\mathbb R}}

\def\ln{\log}
\def\1{I}
\def\imath{\textbf{i}}
\def\jmath{\textbf{j}}
\def\chi{\zeta}

\def\QED{\hfill$\square$\vskip 3mm}

\def\Dp{\displaystyle}
\def\Df{\Dp\frac}

\def\({\left(}
\def\){\right)}

\title{ Minimal Joint Entropy and Order-Preserving Couplings\footnotetext{Research supported in part by the Natural Science Foundation of China (under
grants 11471222, 61973015).}}

\author{Ya-Jing Ma$^*$,\  Feng Wang$^*$,\  Xian-Yuan Wu\footnote{School of Mathematical Sciences, Capital
		Normal University, Beijing, 100048, China. Emails: \texttt{mayajing121@126.com},\ \texttt{wangf@cnu.edu.cn},\ \texttt{wuxy@cnu.edu.cn}},\  Kai-Yuan Cai\footnote{Department of Automatic Control, Beijing University of Aeronautics and Astronautics, Beijing, 100083, China. Email:\ \texttt{kycai@buaa.edu.cn}}
}

\date{}

\begin{document}

 \maketitle
\begin{abstract}
This paper focuses on the extreme-value problem for {\it Shannon entropy} of the joint distribution with given marginals. It is proved that the minimum-entropy coupling must be of {\it order-preserving}, while the maximum-entropy coupling coincides with the {\it independent} one. Note that in this sense, we interpret entropy as a measure of {\it system disorder}.
\end{abstract}
\begin{center}
\begin{minipage}{14cm}
{\bf Keywords and phrases}: Shannon entropy, Mutual information, Isoentropy distribution, Order-preserving coupling.

{\bf Mathematics Subject classification(2020)}: 94A17, 60E15
\end{minipage}
\end{center}
\section{Introduction and statement of the result}
\renewcommand{\theequation}{1.\arabic{equation}}
\setcounter{equation}{0}
The concept of entropy was introduced in thermodynamical and statistical mechanics as a measure of uncertainty or disorganization in a physical system \cite{Bl1,Bl2}. In 1877, L. Boltzmann \cite{Bl2} gave the probabilistic interpretation of
entropy and found the famous formula $S=\kappa \ln W$. Roughly speaking, the entropy is the logarithm of the number of ways in which the physical system can be configured. The second law of thermodynamical says that the entropy of a closed system cannot decrease.

To reveal the physics of information, C. Shannon (1948)\cite{S} introduced the entropy in the communication theory. Let $X$ be a discrete random element with alphabet ${\cal X}$ and probability mass ${\bf p}=\{p(x)=\P(X=x): x\in {\cal X}\}$, the entropy of $X$ (or $\bf p$) is defined by
\be\label{1.1} H(X)=H({\bf p}):=-\sum_{x\in{\cal X}}p(x)\log p(x).\ee Clearly, $H(X)$, which is called the {\it Shannon entropy}, takes its minimum $0$ when $X$ is degenerated and takes its maximum $\log |{\cal X}|$ when $X$ is uniformly distributed. In this sense, entropy is a measure of the uncertainty of a random element.

In the theory of information, the definition of entropy is extended to a pair of random variables as follows. Let $(X,Y)$ be a two-dimensional random vector in ${\cal X}\times{\cal Y}$ with a joint distribution $P$=$\{p(x,y):x\in{\cal X},y\in{\cal Y}\}$, the {\it joint entropy} of $(X,Y)$ (or $P$) is defined by
\be\label{1.2}H(X,Y)=H(P):=-\sum_{x\in{\cal X}}\sum_{y\in{\cal Y}}p(x,y)\log p(x,y).\ee Another important concept on $(X,Y)$ is the {\it mutual information}, which is a measure of the amount of information that one random variable contains about the other. It is defined by
\be\label{1.3}I(X,Y):=\sum_{x\in{\cal X}}\sum_{y\in{\cal Y}}p(x,y)\log {\Df {p(x,y)}{p(x)p(y)}},\ee
where $\{p(x):x\in{\cal X}\}$, $\{p(y):y\in{\cal Y}\}$ are the marginal distributions of $X,Y$. By definition, one has
\be\label{1.4}I(X,Y)=H(X)+H(Y)-H(X,Y).\ee
Note that in some setting, the {\it maximum} of {\it mutual information} is called {\it channel capacity}, which plays a key role in information theory through the famous {\it Shannon's second theorem: Channel Coding Theorem} \cite{S}.

For basic concepts and properties in information theory, readers may refer to \cite{CT} and the references therein.

By (\ref{1.4}), for given marginals $\{p(x):x\in\cal X\}$ and $\{p(y):y\in \cal Y\}$, maximizing $I(X,Y)$ and minimizing $H(X,Y)$ are two sides of a single coin. To infer an unknown joint distribution of two random
variables with given marginals is an old problem in the area of probabilistic inference. As far as we know, the problem may go back at least to Hoeffding \cite{H} and Frechet \cite{F}, who
studied the question of identifying the extremal joint distribution that maximizes (resp., minimizes) their correlation. For more literatures in this area and more applications in pure and applied sciences, readers may refer to \cite{BS,CFR,CGV,DKS,LDKH} etc.

In this paper, we consider the following case of the problem described above. For simplicity, suppose ${\cal X}={\cal Y}=\{1,2,\ldots,n\}, n\geq2$ and ${\bf p}$$=(p_1,p_2,\ldots,p_n)$, ${\bf q}$$=(q_1,q_2,\ldots,q_n)$ be two discrete probability distributions on $\cal X$. First we associate random variables $X$, $Y$ in $\cal X$ to ${\bf p}$ and ${\bf q}$ in some way (see the following Definition~\ref{d1}) respectively, second we seek a {\it minimum-entropy} two-dimensional random vector $(X,Y)$ in ${\cal X}\times{\cal X}$ with marginals $\bf p$ and $\bf q$ (see the optimization problem (\ref{1.6})).

One strategy to solve the problem mentioned above is to calculate the exact value of the minimum entropy $H(X,Y)$. Since, for general case, the corresponding optimization problem is known to be NP-hard, people prefer to give it good estimates. Recently, F. Cicalese etc. \cite{CGV} solved the problem almost perfectly in this respect. Actually, they obtained an efficient
algorithm to find a joint distribution with entropy exceeding the minimum at most by 1 (see Remark~\ref{r2}).

Another strategy to study the problem is to seek the unknown special structure of a minimum-entropy coupling $(X,Y)$. Clearly, in the case when $X$ and $Y$ are {\it independent}, the joint entropy $H(X,Y)$ takes the maximum $H(X)+H(Y)$. That is to say, the {\it independent} structure ( maybe the {\it most} disordered one) of $(X,Y)$ determines the maximum entropy, but what special structure in a coupling $(X,Y)$ will determine the minimum entropy of the two-dimensional random system? The main goal of the present paper is to establish such a structure.

Denote by ${\cal P}_n$ the set of all discrete probability distributions on ${\cal X}$ $=\{1,$ $2,\ldots,n\}$. For each $\bf p\in{\cal P}_n$, let $F_{\bf p}$ be the cumulative distribution function defined by
$$F_{\bf p}(i):=\sum_{k=1}^ip_k,\ 1\leq i\leq n.$$

Recall that a permutation $\sigma$ is a bijective map from $\cal X$$=\{1,2,\ldots,n\}$ into itself. For any given distribution ${\bf p}=(p_1,p_2,\ldots,p_n)\in{\cal P}_n$, define $\sigma{\bf p}:=(p_{\sigma(1)},p_{\sigma(2)},\ldots,p_{\sigma(n)})$. By the definition (\ref{1.1}), one has
\be\label{1.5}H({\bf p})=H(\sigma{\bf p})\ee
holds for any permutation $\sigma$. Based on this fact, in the present paper, we identify all $\sigma{\bf p}$'s with ${\bf p}$ as one distribution on $\cal X$.
To this end, we define an equivalence relation $``\sim"$ in ${\cal P}_n$: for any ${\bf p, p'}\in{\cal P}_n$, ${\bf p}\sim{\bf p'}$ if and only if for some permutation $\sigma$, ${\bf p'}={\sigma} {\bf p}$. By (\ref{1.5}), the entropy is constant for all distributions in an equivalent class. Denote by $\bar{\cal P}_n$ the subset of all ${\bf p}\in{\cal P}_n$ such that $p_1\geq p_2\geq\ldots\geq p_n$. Obviously, $\bar{\cal P}_n$ is an isomorphism of the quotient space ${\cal P}_n/\sim$, we should identify $\bar{\cal P}_n$ with ${\cal P}_n/\sim$ in case of necessity. For each ${\bf p}\in \bar{\cal P}_n$, we call it an {\it isoentropy} distribution,


Now, for isoentropy distributions in $\bar{\cal P}_n$, we give the following definitions.

\bd\label{d1} Given isoentropy distributions ${\bf p,q}\in\bar{\cal P}_n$. Suppose $X$ is a random variable in $\cal X$ and $(X,Y)$ is a two-dimensional random vector in $\cal X\times\cal X$ with joint distribution matrix $P$.
\begin{enumerate}
	\item Random variable $X$ is distributed according to isoentropy distribution ${\bf p}$, if for some permutation $\sigma$, $X$ is distributed according to $\sigma \bf p$.
	
	\item Random vector $(X,Y)$ (or its joint distribution $P$) is called having marginals $\bf p$ and $\bf q$, if for some permutations pair $\sigma,\sigma'$, $P$ has marginals $\sigma \bf p$ and $\sigma'\bf q$.
	
\end{enumerate}
\ed
\br For an isoentropy distribution ${\bf p}\in \bar{\cal P}_n$, if $X$ is distributed according to $\bf p$, then only the entropy of $X$ is determined. But in the classical situation, $X$ being distributed according to some ${\bf p}\in{\cal P}_n$ implies that all digital features (especially including the moments) of $X$ are determined by its distribution $\bf p$.      \er

For any ${\bf p,q}\in {\cal P}_n$, denote by ${\cal C}(\bf p,q)$ the set of all joint distributions with marginals $\bf p,q$.
For any ${\bf p, q}\in \bar{\cal P}_n$, denote by ${\cal C}_e(\bf p,q)$ the set of all joint distribution matrix $P$ with marginals $\bf p,q$ (according to Definition~\ref{d1}, item $2$). For any $P\in {\cal C}_e(\bf p,q)$, with a little abuse of terminology, we call $P$ a {\it coupling} of $\bf p,q.$ 

Now we turn to the following optimization problem: to find a $\hat P\in {{\cal C}_e}(\bf p,q)$, such that
\be\label{1.6}H(\hat P)=\inf_{P\in{{\cal C}_e}(\bf p,q)}H(P).\ee
Note that ${\cal C}_e(\bf p,q)$ forms a compact subset of $\R^{n^2}$, the existence of such a $\hat P$ follows from the continuity of the entropy function $H$.

\br\label{r2} For any ${\bf p,q}\in \bar{\cal P}_n\subset{\cal P}_n$, let ${\bf p}\wedge{\bf q}$ be the distribution with cumulative distribution function $F_{{\bf p}\wedge{\bf q}}=F_{{\bf p}}\wedge F_{{\bf q}}$.
F. Cicalese etc. obtained the following relation in \cite{CGV}
\be\label{1.7}
H({\bf p}\wedge{\bf q})\leq H(\hat P)=\inf_{P\in{{\cal C}_e}(\bf p,q)}H(P)\leq H({\bf p}\wedge{\bf q})+1.\ee
In fact, to get the upper estimate, \cite{CGV} constructed a $P\in {\cal C}_e(\bf p,q)$ from ${\bf p}\wedge{\bf q}$ such that $H(P)\leq H({\bf p}\wedge{\bf q})+1$, but no special structure of that $P$ is worthy of attention.

\er

For any $1\leq k, l\leq n$, denote by $E(k,l)=\(e_{i,j}\)_{n\times n}$ the $n$-th order square matrix such that
$e_{i,i}=1, \ i\neq k,l$, $e_{k,l}=e_{l,k}=1$, and $e_{i,j}=0$ otherwise, let $\cal  E$$:=\{E(k,l):1\leq k,l\leq n\}$. For any $n$-th order probability matrix $P$, let $P'=E(k,l)\cdot P$ (resp. $P\cdot E(k,l)$), then $P'$ is the matrix obtained from $P$ by exchanging the positions of its $k$-th and $l$-th rows (resp. columns). For any ${\bf p,q}\in \bar{\cal P}_n$, we define an equivalent relation $``\sim"$ in ${\cal C}_e({\bf p,q})$: $P,Q\in {\cal C}_e({\bf p,q})$ is called equivalent to each other, write as $P\sim Q$, if for some subsets ${\cal E}^1=\{E^1_i:1\leq i\leq m_0\}, {{\cal E}^2=\{E_j^2:1\leq j\leq n_0\}}\subset {\cal E}$, $m_0,n_0\geq 1$, such that
$$P=\(\prod_{i=1}^{m_0}E^1_i\)\cdot Q\cdot \(\prod_{j=1}^{n_0}E^2_j\).$$ In other words, $P\sim Q$ if and only if $P,Q$ can be obtained from each other by exchanging row and column positions, hence $H(P)=H(Q)$.

Now, for any ${\bf p,q}\in{\cal P}_n$, ${\cal C}(\bf p,q)$ is an isomorphism of the quotient space ${\cal C}_e( {\bar {\bf p},\bar {\bf q}})/\sim$, where ${\bar {\bf p},\bar {\bf q}}\in \bar{\cal P}_n$, $\bar{\bf p}\sim \bf p$, $\bar{\bf q}\sim \bf q$. On account of the fact that
$$\inf_{P\in {\cal C}(\bf p,q)}H(P)=\inf_{P\in {\cal C}_e({\bar {\bf p},\bar {\bf q}})}H(P),$$ the optimization problem (\ref{1.6}) is equivalent to the following original one
\be\label{1.6'}\tilde P: H(\tilde P)=\inf_{P\in{\cal C}(\bf p,q)}H(P).\ee

In order to describe the special structure of a minimum-entropy coupling $\hat P$ given in (\ref{1.6}), we give the following definition.
\bd\label{d2} For any ${\bf p,q}\in \bar{\cal P}_n$, a coupling $P\in {\cal C}_e(\bf p,q)$ is called order-preserving, if $P$ is upper triangular, i.e., for any $1\leq j\leq i\leq n$, $p_{i,j}=0$. In other words, if $(X,Y)$ is distributed according to $P$, then
\be\label{1.8}\P(X\leq Y)=1.\ee
Denote by ${\cal O}(\bf p,q)$ the set of all order-preserving couplings of ${\bf p,q}\in \bar {\cal P}_n$.
\ed

\bp\label{p1}For any $n\geq 2$, and for any $\bf p,q$$\in\bar{\cal P}_n$, ${\cal O}(\bf p,q)\neq \emptyset$.\ep

{\it Proof.} Look $\bf p,q$ as elements in ${\cal P}_n$, let ${\tilde {\bf q}}\in {\cal P}_n$ satisfying ${\tilde {\bf q}}\sim\bf q$ and $\tilde q_1\leq\tilde q_2\leq\ldots\leq \tilde q_n$. It is straightforward to check that, for the cumulative distribution functions $F_{\bf p}$ and $F_{ \tilde {\bf q}}$, one has $F_{\bf p}\geq F_{\tilde{\bf q}}.$ Then by the Strassen's Theorem \cite{ST} on stochastic domination, there exists a coupling $P$ of $\bf p$ and $\tilde {\bf q}$, such that
$$\sum_{i=1}^n\sum_{j=i}^n p_{i,j}=1.$$ This means that $P$ is upper triangular and by Definition~\ref{d1}, $P\in{\cal C}_e(\bf p,q)$ then $\in {\cal O}(\bf p,q)$.\QED

Now, we state our main result as the following.

\bt\label{t} Suppose $n\geq 2$ and $\bf p,q$$\in{\bar{\cal P}}_n$. If $\hat P\in{\cal C}_e(\bf p,q)$ is a solution of the optimization problem (\ref{1.6}), then $\hat P$ is order-preserving. In other words, for any ${\bf p},{\bf q}\in {\cal P}_n$, suppose $\bf p\sim\bar{\bf p}$, $\bf q\sim\bar{\bf q}$ and ${\bar {\bf p}, \bar {\bf q}}\in \bar{\cal P}_n$, if $\tilde P\in{\cal C}(\bf p,q)$ is a solution of the optimization problem (\ref{1.6'}), then there exists $\hat P\in {\cal O}(\bar{\bf p},\bar{\bf q})$ such that $\tilde P\sim\hat P$. \et

\br By Theorem~\ref{t}, the optimization problem (\ref{1.6}) can be simplified as the following
\be\label{1.9}\hat P: H(\hat P)=\inf_{P\in{{\cal O}}(\bf p,q)}H(P).\ee With this simplification, firstly, the corresponding computational complexity is well reduced; secondly, the order-preserving structure may possibly help us to construct the concrete form of $\hat P$.

\er
\vskip 5mm
\section{Local optimization lemmas}
\renewcommand{\theequation}{2.\arabic{equation}}
\setcounter{equation}{0}
Suppose $A=(a_{i,j})_{n\times n}$ is a positive square matrix, i.e. all its entries are nonnegative. We define its entropy by
\be\label{2.1}H(A)=-h(A):=-\sum_{i=1}^n\sum_{j=1}^n a_{i,j}\log a_{i,j}.\ee Denote $C=\sum_{i=1}^n\sum_{j=1}^n a_{i,j}$ and let $P=C^{-1}A$, a probability matrix, then
\be\label{2.2} h(A)=C h(P)+C\log C.\ee

For any $c>0$, define $h_c(x)=x\log x+(c-x)\log(c-x), \ x\in [0,c]$. Before the local optimization lemmas, we give out the following simple property for function $h_c$ without proof.

\bl\label{l0} For any closed interval $[a,b]\subseteq[0,c]$, $\max\{h_c(x):x\in[a,b]\}=h_c(a)\vee h_c(b)$$=h_c(a)\1_{\{a\leq c-b\}}+h_c(b)\1_{\{a>c-b\}}.$\el
\bl\label{l1}For any second order positive square matrix $A=\(a_{i,j}\)_{2\times 2}$. Suppose that $a_{1,1}\vee a_{2,2}\geq a_{1,2}\vee a_{2,1}$, let $b=a_{1,2}\wedge a_{2,1}$. Let $A'=\(a'_{i,j}\)_{2\times 2}$ such that $a'_{i,i}=a_{i,i}+b,\ i=1,2$, $a'_{i,j}=a_{i,j}-b,\ i\neq j.$ Then $H(A)\geq H(A')$. Furthermore, if $b>0$, then $H(A)>H(A')$.\el

{\it Proof.} Without lose of generality, assume $a_{1,1}\geq a_{2,1}\geq a_{1,2}$. Note that in this case, one has $b=a_{1,2}$ and $a'_{1,2}=0$. For any $x\in [0,b]$, define
$$A(x)=\(\ba{ll}a_{1,1}+x &a_{1,2}-x\\[5mm]
a_{2,1}-x & a_{2,2}+x\ea\).$$
Let $c_1=a_{1,1}+a_{2,1}$, $c_2=a_{1,2}+a_{2,2}$, then $h(A(x))=h_{c_1}(a_{1,1}+x)+h_{c_2}(a_{1,2}-x)$.

By Lemma~\ref{l0}, and the assumption given above, $\max\{h_{c_1}(a_{1,1}+x):x\in[0,b]\}=h_{c_1}(a_{1,1}+b)$, $\max\{h_{c_2}(a_{1,2}-x):x\in[0,b]\}=h_{c_2}(0).$ Thus $\max\{h(A(x)):x\in[0,b]\}=h(A(b))$ and $h(A)=h(A(0))\leq h(A(b))=h(A')$, then by definition $H(A)\geq H(A')$.

In the case of $b=a_{1,2}>0$, one has $a_{11}>a_{2,1}-a_{1,2}$, then by Lemma~\ref{l0}, $h_{c_1}(a_{1,1})<h_{c_1}(a_{1,1}+a_{1,2})$, this implies $h(A)<h(A')$ and $H(A)>H(A')$.\QED

\bl\label{l2}For any second order positive square matrix $A=\(a_{i,j}\)_{2\times 2}$. Suppose that $a_{1,1}+a_{1,2}\geq a_{2,1}+a_{2,2}, a_{1,1}+a_{2,1}\geq a_{1,2}+a_{2,2}$ and $a_{1,1}+a_{1,2}\geq a_{1,1}+a_{2,1}$. Let $b=a_{1,2}\wedge a_{2,1}$, define $A'$ as in Lemma~\ref{l1}, then $H(A)\geq H(A')$.\el
{\it Proof.} By relation (\ref{2.2}), without lose of generality, assume $A$ be a probability matrix. Rewrite by $P=\(p_{i,j}\)_{2\times 2}$ the probability matrix, and write ${\bf p}=(p_1,p_2)=(p,1-p)$, ${\bf q}=(q_1,q_2)=(q,1-q)$ as its marginals, i.e. $p_{1,1}+p_{1,2}=p, \ p_{2,1}+p_{2,2}=1-p$ and $p_{1,1}+p_{2,1}=q$, $p_{1,2}+p_{2,2}=1-q$. Note that in this case one has $p\geq q\geq 0.5$.

For any $x\in [0,1-p]$, let
$$P(x)=\(\ba{ll} q-x &(p-q)+x\\[5mm]
x & (1-p)-x\ea\).$$
Clearly, $P(x)$ is a joint probability distribution matrix with marginals $\bf p,q$ and $P(p_{2,1})=P$, $P(0)=P'$. Particularly, when $x$ run over the interval $[0,1-p]$, $P(x)$ ran over all couplings of $\bf p,q$.

Now $h(P(x))=h_p(q-x)+h_{1-p}(x)$. It follows immediately from Lemma~\ref{l0} that $h_{1-p}(x)$ and $h_p(q-x)$ take their maximums simultaneously when $x=0$, then $h(P)=h(P(p_{2,1}))\leq h(P(0))=h(P')$ and $H(P)\geq H(P')$.\QED

\br From the statements of Lemmas~\ref{l1} and \ref{l2}, for any second order positive square matrix $A$, one may easily to construct the positive matrix $A'$, which possesses the same row and column summations as $A$, such that $H(A)\geq H(A')$.\er

Now, as a direct consequence of Lemmas~\ref{l1} and \ref{l2}, we obtain the following local optimization theorem.

\bt\label{t1} Suppose ${\bf p,q}\in\bar {\cal P}_n$, $n\geq 2$, and $P\in{\cal C}_e(\bf p,q)$. For any $1\leq i<k\leq n$, $1\leq j<l\leq n$, let
$$A=\(\ba{ll}p_{i,j}& p_{i,l}\\[5mm]
p_{k,j}&p_{k,l}\ea\)$$ be the second order submatrix of $P$. Let $A'$ be given as in Lemma~\ref{l1} or Lemma~\ref{l2} and let $P'$ be the matrix obtained from $P$ by $A'$ taking the place of $A$. Then $P'\in {\cal C}_e(\bf p,q)$ and $H(P)\geq H(P')$.
\et

Finally, as a consequence of Lemma~\ref{l2}, the optimization problem (\ref{1.6}) for $n=2$ is solved as the following.

\bt\label{t2}For any ${\bf p,q}\in \bar{\cal P}_2$, suppose that ${\bf p}=(p,1-p)$, ${\bf q}=(q,1-q)$ and $p\geq q$. Let
$$\hat P=\(\ba{ll}q &p-q\\[5mm]
0& 1-p\ea\),$$ then $$H(\hat P)=\inf_{P\in{\cal C}_e(\bf p,q)}H(P)=-[q\log q+(1-p)\log(1-p)+(p-q)\log(p-q)].$$
\et

\vskip 5mm
\section{Proof of Theorem~\ref{t}}
\renewcommand{\theequation}{3.\arabic{equation}}
\setcounter{equation}{0}

In this section, we give proof to Theorem~\ref{t}. Actually, the strategy of the proof is to use the local optimization Lemma~\ref{l1} repeatedly. First of all, we have the following lemma.

\bl\label{l4} Let $A=\(a_{i,j}\)_{n\times n},\ n\geq 2$ be a $n$-th order positive square  matrix. Suppose that \be\label{3.1}a_{1,1}=\max_{1\leq i,j\leq n}a_{i,j},\ \sum_{j=1}^na_{1,j}\geq\sum_{i=1}^na_{i,1}\ee
$$\({\rm resp.}
\ a_{n,n}=\max_{1\leq i,j\leq n}a_{i,j},\  \sum_{j=1}^na_{n,j}\leq\sum_{i=1}^na_{i,n}
\).$$
Then, by using the local optimization procedure developed in Lemma~\ref{l1} and Theorem~\ref{t1} at most $2(n-1)$ times, we finally transform $A$ to $A'=\(a'_{i,j}\)_{n\times n}$ such that $H(A)\geq H(A')$,
\be\label{3.2}\sum_{i=1}^n a_{i,j}=\sum_{i=1}^n a'_{i,j},\ \sum_{j=1}^n a_{i,j}=\sum_{j=1}^n a'_{i,j},\ 1\leq i,j\leq n,\ {\rm and}\ee
\vskip-5mm
\be\label{3.3}a'_{1,1}=\sum_{i=1}^na_{i,1},\ \ a'_{i,1}=0, \ 2\leq i\leq n\ee $$\({\rm resp.}\ a'_{n,n}=\sum_{j=1}^na_{n,j},\ \ a'_{n,j}=0, \ 1\leq j\leq n-1\).$$ Furthermore, $H(A)=H(A')$ if and only if $a'_{1,1}=a_{1,1}$ $\({\rm resp.}\ a'_{n,n}=a_{n,n}\).$ \el
{\it Proof.} Let $I=\{i:a_{i,1}>0,\ 2\leq i\leq n\},$ $J=\{j:a_{1,j}>0, \ 2\leq j\leq n\}$, denote by $|I|, |J|$ the cardinalities of $I$ and $J$. Without lose of generality, we assume $i_0\in I,\ j_0\in J$ satisfies
$$a_{i_0,1}=\min_{i\in I}a_{i,1}\wedge \min_{j\in J}a_{1,j},\ a_{1,j_0}=\min_{j\in J}a_{1,j}.$$ Write $b=a_{i_0,1}$, we renew $A$ by changing the second order submatrix
$$\(\ba{ll} a_{1,1} & a_{1,j_0}\\[5mm]
a_{i_0,1}&a_{i_0,j_0}\ea\) \ \rm{to}\ \(\ba{ll} a_{1,1}+b & a_{1,j_0}-b\\[5mm]
a_{i_0,1}-b&a_{i_0,j_0}+b\ea\).$$ Then, the renewed $A$ still satisfies the conditions of the Lemma~\ref{l1} and then by this lemma, its entropy $H(A)$ is decreased strictly. Note that after using the local optimization procedure once, for the renewed matrix $A$, the number $|I|+|J|$ decreases by $1$.

Repeat the above procedure until $I=\emptyset$, write $A'$ as the final renewing of $A$, thus $A'$ is obtained as required.\QED

In the situation given as in the statement of Lemma~\ref{l4}, we denote $L_n$ the corresponding composite optimization procedure, and write $A'=L_n(A)$.

Now, we finish the proof of Theorem~\ref{t} by proving the following theorem.

\bt\label{t3} Given ${\bf p,q}\in\bar {\cal P}_n,\ n\geq 2$. For any $P\in {\cal C}_e(\bf p,q)$, there exists $P'\in{\cal O}(\bf p,q)$ such that $H(P)\geq H(P')$. Furthermore, if $P$ is not equivalent to any $Q\in{\cal O}({\bf p,q})$, then there exists a $P'\in {\cal O}({\bf p,q})$ such that $H(P)>H(P')$.\et

{\it Proof.} We prove the theorem by induction on $n$. For $n=2$, Theorem~\ref{t3} follows from Lemma~\ref{l1} immediately.

For general $n>2$ and $P=\(p_{i,j}\)_{n\times n}\in{\cal C}_e(\bf p,q)$, suppose $p_{i_0,j_0}=\Dp\max_{1\leq i,j\leq n}p_{i,j}$.
In the case of $$\sum_{j=1}^np_{i_0,j}\geq \sum_{i=1}^np_{i,j_0}\ \({\rm resp.}\ \sum_{j=1}^np_{i_0,j}\leq \sum_{i=1}^np_{i,j_0} \), $$ let $A=E(1,i_0)\cdot P\cdot E(1,j_0)$ (resp. $E(n,i_0)\cdot P\cdot E(n,j_0)$), where $E(1,i_0)$, $E(1,j_0)$, $E(n,i_0)$, $E(n,j_0)\in {\cal E}$. Then $A\sim P$ and satisfies the condition (\ref{3.1}). By Lemma~\ref{l4}, equations (\ref{3.2}) and (\ref{3.3}) hold for $A'=L_n(A)$, where $L_n$ is the composite optimization procedure introduced before the statement of Theorem~\ref{t3}.

Note that by (\ref{3.2}), $A, A'$ are all probability matrixes in ${\cal C}_e(\bf p,q)$, and $H(P)=H(A)\geq H(A')$.

Now $A'$ has the form
\be\label{3.4}A'=\(\ba{lllll}a'_{1,1}&a'_{1,2}&\cdots& a'_{1,n-1}&a'_{1,n}\\
0&a'_{2,2}&\cdots &a'_{2,n-1}&a'_{2,n}\\
\vdots&\vdots&\ddots&\vdots&\vdots\\
0&a'_{n-1,2}&\cdots &a'_{n-1,n-1} &a'_{n-1,n}\\
0&a'_{n,2}&\cdots&a'_{n,n-1}&a'_{n,n}\ea
\)\ee$$ \({\rm resp.}\(\ba{lllll}a'_{1,1}&a'_{1,2}&\cdots &a'_{1,n-1}&a'_{1,n}\\
a'_{2,1}&a'_{2,2}&\cdots &a'_{2,n-1}&a'_{2,n}\\
\vdots&\vdots&\ddots&\vdots&\vdots\\
a'_{n-1,1}&a'_{n-1,2}&\cdots &a'_{n-1,n-1}&a'_{n-1,n}\\
0&0&\cdots&0&a'_{n,n}\ea
\) \).$$
Consider the following $(n-1)$-th order submatrix $\bar A=\(\bar a_{i,j}\)_{(n-1)\times(n-1)}$ of $A'$,
$$\bar A=\(\ba{llll}a'_{2,2}&\cdots &a'_{2,n-1}&a'_{2,n}\\
\vdots&\ddots&\vdots&\vdots\\
a'_{n-1,2}&\cdots &a'_{n-1,n-1} &a'_{n-1,n}\\
a'_{n,2}&\cdots&a'_{n,n-1}&a'_{n,n}\ea\)$$
$$\({\rm resp.}\(\ba{llll}a'_{1,1}&a'_{1,2}&\cdots &a'_{1,n-1}\\
a'_{2,1}&a'_{2,2}&\cdots &a'_{2,n-1}\\
\vdots&\vdots&\ddots&\vdots\\
a'_{n-1,1}&a'_{n-1,2}&\cdots &a'_{n-1,n-1}\ea\) \),
$$
suppose $a'_{i_1,j_1}$ be the entry of $\bar A$ such that $a'_{i_1,j_1}=\Dp\max_{1\leq i,j\leq n-1}\bar a_{i,j}$.

For simplicity, we only treat the first case of $\bar A$ (the resp. case in the brackets can be treated similarly), of course, in this case one has $2\leq i_1,j_1\leq n.$

In the subcase when
$$\sum_{i=2}^na'_{i,j_1}\leq\sum_{j=2}^na'_{i_1,j}\ \ \({\rm resp.}\ \sum_{i=2}^na'_{i,j_1}\geq\sum_{j=2}^na'_{i_1,j}\),$$ let $A''=(a''_{i,j})_{n\times n}=E(2,i_1)\cdot A'\cdot E(2,j_1)$ (resp. $E(n,i_1)\cdot A'\cdot E(n,j_1)$). Clearly, one has $$A''\in {\cal C}_e({\bf p,q}), \ H(P)\geq H(A')=H(A'').$$ We point out here that, $A''$ has the same form of $A'$ as given in equation (\ref{3.4}), and $\bar{\bar A}:=\(a''_{i,j}:2\leq i,j\leq n\)$ as the $(n-1)$-th order submatrix of $A''$ satisfies the condition of Lemma~\ref{l4} for $n-1$.

At this moment, we have finished the {\it first} step of optimization of $P$ by transforming $P$ to $A''$. Actually, we are standing at the position to begin the {\it second} step by using Lemma~\ref{l4} to $\bar{\bar A}$ for $n-1$.

Repeat the above procedure for $n-1$, $n-2$, $\cdots$, and finally for $2$, $A''$ is then transformed to $A''''$, $A^{(6)}$, and finally to $P'=A^{(2(n-1))}$. By our setting, $P'\in {\cal C}_e({\bf p,q})$ and is upper triangular, i.e. $P'\in{\cal O}({\bf p,q})$.

If $P$ is not equivalent to any $Q\in{\cal O}({\bf p,q})$, then at least one optimization step carried out above is concrete, i.e., when Lemma~\ref{l1} is used, the corresponding $b$ is strictly positive, then by Lemma~\ref{l1} and the followed Theorem~\ref{t1}, $H(P)>H(P')$.\QED


\end{document}